\documentstyle[12pt]{article}
\begin{document}
\makeatletter
\@addtoreset{equation}{section}
\makeatother
\renewcommand{\theequation}{\thesection.\arabic{equation}}
\baselineskip 15pt

\title{\bf On a recent proof of nonlocality without inequalities}

\author{GianCarlo Ghirardi\footnote{e-mail: ghirardi@ts.infn.it}\\ {\small Department of  Physics of the University of Trieste, and}\\
{\small the Abdus Salam International Centre for Theoretical Physics,
Trieste, Italy.}}

\date{}

\maketitle

\begin{abstract}
Recently a quite stimulating paper \cite{vedral} dealing with the possibility of exploiting the nonlocal aspects of a superposition of states of a single photon appeared. We regard as greatly relevant the results which have been obtained. However we think that the presentation of the matter and the way to derive the conclusion are not fully satisfactory  and do not put the necessary emphasis on some subtle basic aspects like locality and realism.  In view of its interest we consider it useful to reconsider the line of reasoning of ref.[1] and to derive once more its results by following a procedure which seems to us more lucid and which makes fully clear the role of the various conceptual aspects of the treatment. We hope that our analysis will contribute to clarify and to deepen the interesting results of ref.[1].
\end{abstract}

\section{Introductory remarks}
Few months ago, a quite interesting paper appeared \cite{vedral} in which a (practically) all versus nothing violation of (using the authors' terminology) {\it local realism} has been derived.  The specific interest of the paper stems from its dealing with a single photon in a superposition of different locations. We will not discuss this important point here while we will reconsider the derivation of the result. Let us make clear from the very beginning that, as we will show in what follows, the conclusions of the paper are  correct. However we believe that the details of the procedure to reach such conclusions are not fully satisfactory. In particular, as we will stress below, ref.[1] has not made sufficiently clear the crucial interplay between the requests of locality and those of realism.  Even more, the very fundamental role of the locality assumption and the way in which one has to use it  to derive the result, are not made completely explicit. Accordingly, a critical reconsideration of the basic points on which a conceptually rigorous derivation of the result  must rest is appropriate. 

In this paper, even though we will argue along lines similar to those of ref.[1], we will present an alternative way to reach the same conclusion. In particular, by following the approach we have used recently to deal with similar situations \cite{Marinatto}, we will make a systematic use of set theoretical arguments concerning the space of the states which one uses to characterize in the most accurate possible way the situation of individual physical systems. The treatment, which is entirely and solely based on the locality request, will be extremely general and will hold true for a large class of theories, including  quantum mechanics itself, genuine hidden variable theories or even theories like Bohmian Mechanics in which both controllable and uncontrollable variables enter into play. We will not need to take into account conditional probabilities and we will avoid the tacit use of counterfactual arguments which underlies the procedure of the original paper. In brief, we consider our approach as yielding a logically more consistent, illuminating and simple derivation of the important result of ref.[1]. Our considerations can be seen as supplementing, clarifying and putting on a firm basis  the procedure of the paper under discussion.

\section{ Some general specifications}
As already stated,  our main purpose is to make fully clear the crucial interplay between locality and realism in the derivation. To reach this goal  we will   follow a  line of thought which parallels, with appropriate changes and a richer argumentation, the reasoning leading from Einstein's incompleteness argument to Bell's inequality \cite{Bell0} or the one which has to be followed to get the result \cite{GHZ} of GHZ. 

Let us begin by recalling the conceptually more relevant points of Bell's derivation of his famous inequality. To make our argument simpler we will make reference, as usual, to the familiar case of the singlet state of two spin $1/2$ particles or to the analogous state of two entangled photons characterizing Aspect's experiments.  The crucial issue is to make crystal clear the status and the interplay of the  assumptions of locality and realism  (equivalently, of the existence of possesed properties independently of any measurement procedure). Even more, what has to be stressed strongly is that only the request of locality enters into play and that any statement concerning elements of physical reality can be derived from it\footnote{This being the situation, the reader will understand the reasons for which we do not consider appropriate to speak of {\it local realism}.}. 

Actually, it is extremely important to keep clearly in mind that J. Bell,  to derive his result has never made resort to any  request of realism or of the existence of hidden variables. All his argument rests on  the sole request of locality. This point has been made  by him various times. As an example I will mention one sentence of ref.\cite{Bell1}: {\it It is remarkably difficult to get this point across, that determinism is not a presupposition of the analysis}. He also felt the necessity of adding  in a footnote: {\it My own first paper on this subject starts with a summary of the EPR argument from locality to deterministic hidden variables.  But the commentators have almost universally reported that it begins with deterministic hidden variables.} 

Let us mention the reasons for which this statement is fully appropriate.  Bell's argument holds for a completely general theory such that the maximal specification which is in principle possible for the states of an individual physical system uniquely determines all probabilities of the outcomes of single and multiple measurements of all conceivable observables. Within such a framework one puts forward the locality request ($B-Loc$), i.e. one assumes that, in the case of space-like measurement procedures, the probability of getting an outcome does not depend on the fact that  others measurements are performed or not and, if performed, by the outcomes which are obtained. 

To be more specific  if - in the case of two spin $1/2$ particles in the singlet state - we denote as {\bf a} and {\bf b} the settings of the  measuring apparatuses, as A and B the two possible outcomes  (which can take only the values $+1$ and $-1$) and as $\lambda$ the  maximal specification which is in principle possible for the state of an individual physical system\footnote{Note that, as already mentioned, here $\lambda$ may stand for the wavefunction $\psi$ if one is dealing with a theory as quantum mechanics, for the hidden variables of a genuine HVTheory and for the wavefunction $\psi$ and the particle positions of Bohmian Mechanics.}, the locality request relating the joint and single measurements reads:

\begin{equation}
P(A,B|{\bf a},{\bf b};\lambda)=P(A|{\bf a},*;\lambda)\cdot P(B|*,{\bf b};\lambda),
\end{equation}

\noindent where the star denotes that the corresponding measurement is not performed. Then, the fact that $P(A,A|{\bf a},{\bf a};\lambda)=0$ and Eq.(2.1) imply that either $P(A|{\bf a},*;\lambda)=0$ or $P(A|*,{\bf a};\lambda)=0$. In the first case, since $P(A|{\bf a},*;\lambda)+P(\neg A|{\bf a},*;\lambda)=1$ one gets $P(\neg A|{\bf a},*;\lambda)=1$.  Proceeding in this way one proves that all single probabilities and, as a consequence of Eq.(2.1), all probabilities, take only either the value 0 or the value 1. In short: {\it the locality request} and {\it the perfect quantum correlations} {\bf imply} {\it determinism}, i.e., the existence of properties possessed before any test is made. Analogous considerations can be worked out for  the GHZ case.

What we will prove here is that an argument which, from a logical point of view, parallels, with the due changes and refinements, the one we have just mentioned can be developed for the case under consideration. Our way of tackling the problem will make much more clear the role of the locality assumption and will replace the, in our opinion, not compelling and obscure authors' statement that the properties (1),(2) and (3) of ref.[1] imply that the local outcomes have predefined values before any measurement.

There is a second point which has to be mentioned, i.e.  is  that, within any conceivable theoretical scheme of the kind we mentioned above, after having assumed (or proved, as we will do) that objective properties are possesed by an individual system,  to proceed  one must also make an assumption which is quite obvious but which nevertheless is necessary and it is usually referred to as  {\it faithful measurement}, amounting to state that, when a value is possessed,  performing a measurement of the corresponding observable  one gets  with certainty such a value as the outcome. This point, as trivial as it can appear, is necessary and is usually made fully explicit in discussions on locality and hidden variable theories (see for instance, ref.\cite{Redhead}).

 A final remark concerning ref.[1].  The argument under point (iii) of the paper does not seem fully clean to us since in the previous analysis the authors have proved, e.g., that $\sigma_{X}^{(1)}=\sigma_{X}^{(N)}$ under  certain conditions for the outcomes of a set of measurements of the Z-components  while to prove that $\sigma_{X}^{(2)}=\sigma_{X}^{(N)}$ they condition under a different set of outcomes. It is then  not fully clear how one can argue from the above that  $\sigma_{X}^{(1)}=\sigma_{X}^{(2)}$. 

\section{An alternative derivation of the result of ref.[1]}
As already anticipated we will now go through a proof of the interesting result of ref.[1] by resorting to a different procedure which, in our opinion, makes more clear  the logic of the argument. Let us begin by specifying the quantum and the hidden variables aspects of the problem and then deriving  the conclusion.

\subsection{The quantum state and its implications}
The authors take into account a quantum system of one photon which can be distributed over  N sites (representing different spatial modes for the photon) which may be occupied or empty. Formally, this amounts to consider the Hilbert space $ H= H^{(1)}\otimes  H^{(2)}\otimes ... H^{(N)} $, where each of the $ H^{(i)}$ is a 2-dimensional Hilbert space. In the space $ H^{(i)}$ we introduce a basis $\{|0\rangle_{i},|1\rangle_{i}\}$ which, for simplicity, we identify with the eigenstates belonging to the eigenvalues $+1$ and $-1$, respectively, of the observables $\sigma_{Z}^{(i)}$. The overall state is:
\begin{equation}
|\Psi_{W}\rangle_{N}=\frac{1}{\sqrt{N}}[|1,0,...,0\rangle+|0,1,...,0\rangle+...+|0,0,...,1\rangle].
\end{equation}

\noindent From the form of the statevector, the authors of ref.[1] identify a set of  quantum probabilties (their Eq.(2))  and a set of  conditional quantum probabilities (their equation (3)) taking the value 1. We consider it more appropriate for our purposes to make reference to perfectly analogous relations which identify vanishing quantum probabilities\footnote{The notation that we use parallels exactly the one we have used above in discussing Bell's inequality, and takes into account the fact that, for the moment, we are in a genuinely quantum context.} :
\begin{equation}
P(\sigma_{Z}^{(i)}=-1,\sigma_{Z}^{(j)}=-1;\;\psi_{WN})=0, \forall i,j;\;\;i\neq j.
\end{equation}
\noindent and:
\begin{eqnarray}
P(\sigma_{X}^{(r)}  =  +1& , &\sigma_{X}^{(s)}=-1\;, \;\sigma_{Z}^{(k\neq r,s)}=+1\;;\;\psi_{WN})=0 \nonumber \\
P(\sigma_{X}^{(r)}  = -1 & , &\sigma_{X}^{(s)}=+1\;,\; \sigma_{Z}^{(k\neq r,s)}=+1\;;\;\psi_{WN})=0.
\end{eqnarray}

\subsection{The new general argument}
Following the attitude of Einstein and Bell described above, we pass now to specify the  general theory we will deal with. Let us take into account  a hypothetical theory accounting for the process we are interested in such that the maximal specification which is in principle possible for the states of an individual physical system, which we will express by specifying a variable $\lambda$, uniquely determines all probabilities of the outcomes of single and multiple measurements of all conceivable observables. Actually,  since we are only interested in the probabilities of single and joint measurements of the Z and X components of the spin-variables ($\equiv$ spatial photon modes) we will confine our considerations to the probabilities which the theory attaches to the outcomes of measurements for these variables, i.e. to expressions of the type:

\begin{equation}
P(\sigma_{L}^{(l)}  = a  , \sigma_{J}^{(j)}=b,..., \sigma_{K}^{(k)}=z;\lambda),
\end{equation}

\noindent where $L,J,...,K$ take the values $Z$ or $X$, and $a,b,...,z$ the values $+1$ or $-1$. The number of observables appearing in the probability lies between $1$ and $N$.

At this point we can impose to our general theory to satisfy the analogous of  Bell's locality condition, i.e. that, since the measurements concerning whether the $N$ sites are occupied or empty  take place at the same time, and therefore are space-like with each other, all probabilities  factorize in products of individual probabilities for the indicated outcomes. In formal terms, we require that:

\begin{equation}
P(\sigma_{L}^{(l)}  = a  , \sigma_{J}^{(j)}=b,..., \sigma_{K}^{(k);\lambda}=z;\lambda)=P(\sigma_{L}^{(l)}  = a;\lambda)\cdot P(\sigma_{J}^{(j)}  = b;\lambda) ... \cdot P(\sigma_{K}^{(k)}  = z;\lambda).
\end{equation}

We stress that this is the appropriate way to impose the locality request to our completely general hypothetical theory. Conditions like the one written above are not introduced and discussed in ref.[1] and precisely for this reason the treatment turns out to be not fully satisfactory.

We can go on as follows:
\begin{itemize}

\item Use of the locality condition for Eq.(3.2) yields

\begin{equation}
P(\sigma_{Z}^{(i)}=-1;\lambda)\cdot P(\sigma_{Z}^{(j)}=-1;\lambda)=0,\;\; \forall i\neq j.
\end{equation}

\noindent Now, if $P(\sigma_{Z}^{(i)}=-1;\lambda)=0$ then, due to the fact that $P(\sigma_{Z}^{(i)}=-1;\lambda)+P(\sigma_{Z}^{(i)}=+1;\lambda)=1$ we have that $P(\sigma_{Z}^{(i)}=+1;\lambda)=1$. In a similar way, if $P(\sigma_{Z}^{(i)}=-1;\lambda)\neq 0$, then it is $P(\sigma_{Z}^{(j)}=-1;\lambda)$ which must vanish, implying $P(\sigma_{Z}^{(j)}=+1;\lambda)=1$.

Since the argument holds for arbitrary $i$ and $j$ we can draw our first conclusion: in any local theory satisfying condition (3.2), the specification of the state of the system implies that all the individual probabilities of outcomes for the measurement of the observables $\sigma_{Z}^{(m)}$ can take only either the value 1 or 0. Locality and the quantum correlations for such measurements imply determinism for them, i.e., the values of the corresponding observables are predetermined and do not depend on their being measured or not.

\item Let us proceed. At this point we must enrich the argument with respect to the one used by Bell. We consider the set $\Lambda$ of all possible values of the variable $\lambda$ specifying the state and we define the subset $\Lambda_{k}$ of $\Lambda$ by stating that, if $\lambda \in \Lambda_{k}$ then $\sigma_{Z}^{(k)}(\lambda)=-1$. Note that the sets $\Lambda_{k}$ are disjoint (once more as a consequence of condition (3.2)) and that their union is the whole set $\Lambda$:

\begin{equation}
\Lambda_{m}\cap \Lambda_{n}=\emptyset;\;\; \cup_{m}\Lambda_{m}=\Lambda.
\end {equation}.

Let us confine  our attention to a $\lambda\in\Lambda_{k}$. From the previous argument proving that the assignement of the state determines the value of all $Z$-components and that it cannot happen that two such observables have the value $-1$, we can directly infer that for $\lambda\in\Lambda_{k}$ all $\sigma_{Z}^{(j)}$ for $j\neq k$ take the value $+1$. Now we can pass to take into account the conditions (3.3). The first of the conditions, using the locality request implies:

\begin{equation}
P(\sigma_{X}^{(s)}=+1)\cdot P(\sigma_{X}^{(k)}=-1)\cdot P(\sigma_{Z}^{(i)}=+1,\sigma_{Z}^{(j)}=+1,...,\sigma_{Z}^{(y)}=+1)=0,\;\;\forall i,j,...,y\neq s,k.
\end{equation}

\noindent However, for $\lambda\in\Lambda_{k}$ we know that the last probability (the third factor at the l.h.s.  of Eq.(3.8)) takes the value 1, so that, for $\lambda\in\Lambda_{k}$:

\begin{equation}
P(\sigma_{X}^{(s)}=+1)\cdot P(\sigma_{X}^{(k)}=-1)=0.
\end{equation}

Just in the same way, using the second of Eqs. (3.3) one shows that:
 
 \begin{equation}
P(\sigma_{X}^{(s)}=-1)\cdot P(\sigma_{X}^{(k)}=+1)=0.
\end{equation}

Now, consider Eq.(3.9) and suppose that $P(\sigma_{X}^{(s)}=+1)=0$, which implies $P(\sigma_{X}^{(s)}=-1)=1$. This, when one takes into account Eq.(3.10), implies that $P(\sigma_{X}^{(k)}=+1)=0$ which in turn implies $P(\sigma_{X}^{(k)}=+1)=1$. Arguing in this way we show that, for 
$\lambda\in\Lambda_{k}$ and for any $s$, $\sigma_{X}^{(s)}(\lambda)$ and $\sigma_{X}^{(k)}(\lambda)$
have  precise values (i.e. are predetermined by the assignment of $\lambda$) and such values coincide.

\item The final step is obvious. Since in the previous argument $k$ is completely arbitrary, the sets $\Lambda_{k}$ are disjoint and their union is $\Lambda$, the above conclusion of the coincidence of the possessed values of the X-components holds for all values of $s$ and $k$: $\sigma_{X}^{(s)}(\lambda)=\sigma_{X}^{(k)}(\lambda),\;\;\forall s,k $.

\end{itemize} 

The conclusion is straightforward: any Local Theory compatible with the two probabilistic requests which have been listed in Subsection 3.1 (which straightforwardly follow from the form (3.1) of the quantum state) implies, when the {\it faithful measurement} assumption is made,  that subjecting all sites (systems) to a measurement of $\sigma_{X}^{(r)}$ for all $r$'s, one gets the same outcome for all of them.

A final remark. If the variables specifying the state of the individual physical systems of the hypothetical theoretical scheme we are considering are controllable (as in the case of quantum mechanics),  such a theory, since it reproduces the predictions (3.2) and (3.3) of the standard theory, must attach directly (i.e. without any need of averaging over uncontrollable variables) zero probability to the occurrence of two different outcomes when the X-components are measured. On the other hand, if the variables specifying the state are uncontrollable, averaging over them by using their appropriate probability density distribution $\rho(\lambda)$  cannot change the above conclusions. Accordingly, we have rederived the result of ref.[1] for any conceivable theory of the kind we have made precise at the beginning of Sect. 3.2 and satisfying the locality condition.

\section {The quantum predictions }

It is now  straightforward to compare this conclusion with the predictions of quantum mechanics. In the case of the state $|\Psi_{W}\rangle_{N}$, quantum mechanics attaches the value $\frac{N}{2^{N-1}}$ to the probability of getting  equal outcomes for the measurements of the $X$ components of all particles, to be compared with the probability 1 attached to such a set of outcomes by any of the theories we have taken into account. Incidentally this conclusion shows that quantum mechanics itself must violate the locality request, since it implies relations (3.2) and (3.3.) but it also implies that one can get, with appreciable probability, different outcomes in the measurement of two observables $\sigma_{X}^{(m)}$ for different $m's$. Obviously, the quantum violation of the locality request can also be exhibited in a very direct and elementary way\footnote{For the interested reader we mention that actually quantum mechanics violates the locality request by violating, in A. Shimony's terminology, {\it Outcome Independence}, i.e. in it, performing a measurement at a given site can change the probability of a space-like separated measurement according to the outcome which is obtained in the first measurement. The theory does not violate {\it Parameter Independence}, i.e., the probability of a far away result does not change if one performs a measurement but he does not read the outcome. As well known, the logical conjunction of the requests of {\it Outcome Independence} and of {\it Parameter Independence} implies and is implied by Bell's locality condition.}.

Moreover, as appropriately remarked in ref.[1] the value $\frac{N}{2^{N-1}}$ tends to zero for increasing $N$ and this is what makes the derived violation of locality practically of the all versus nothing type.

\section {Conclusions}
We think that our analysis should have made clear that, for a quantum system  described by the pure state $|\Psi_{W}\rangle_{N}$, the sole request of locality implies determinism (i.e. preassigned values of the observables) and also implies that all quantum predictions cannot be reproduced. The logic of our argument is perfectly clean and matches, with the appropriate changes, the one which  characterizes the work of J. Bell.

We have presented what we consider a  more satisfactory and illuminating derivation of the  results of ref.[1], results which, we stress once more, are, at any rate, correct. Our work  supplements and makes more precise the derivation of ref.[1],  contributing in this way to put its extremely interesting and relevant conclusion  on firm logical grounds.

\end{document}